\documentclass[]{spie}  

\usepackage{amsmath,amsfonts,amssymb}
\usepackage{graphicx}
\usepackage{siunitx}
\usepackage[colorlinks=true, allcolors=blue]{hyperref}

\title{Infrared photometry with InGaAs detectors: \\First light with SPECULOOS}

\author[a,b]{Peter P. Pedersen}
\author[a,b]{Didier Queloz}
\author[c]{Lionel Garcia}
\author[b]{Yannick Schacke}
\author[d]{Laetitia Delrez}
\author[e]{Brice-Olivier Demory}
\author[f]{Elsa Ducrot}
\author[g]{Georgina Dransfield}
\author[d]{Michael Gillon}
\author[a]{Matthew J. Hooton}
\author[a]{Clàudia Janó-Muñoz}
\author[d]{Emmanuël Jehin}
\author[g]{Daniel Sebastian}
\author[d]{Mathilde Timmermans}
\author[a]{Samantha Thompson}
\author[g]{Amaury H.M. J. Triaud}
\author[h]{Julien de Wit}
\author[d]{Sebastián Zúñiga-Fernández}

\affil[a]{Cavendish Laboratory, University of Cambridge, Cambridge, United Kingdom}
\affil[b]{Department of Physics, ETH Zurich, Zurich, Switzerland}
\affil[c]{Center for Computational Astrophysics, Flatiron Institute, New York, NY, USA}
\affil[d]{Astrobiology Research Unit, Université de Liège, Liège, Belgium}
\affil[e]{Center for Space and Habitability, University of Bern, Bern, Switzerland}
\affil[f]{LESIA, Observatoire de Paris, CNRS, Université Paris Diderot, Université Pierre et Marie Curie, Meudon, France}
\affil[g]{School of Physics \& Astronomy, University of Birmingham, Birmingham, United Kingdom}
\affil[h]{Department of Earth, Atmospheric and Planetary Science, Massachusetts Institute of Technology, Cambridge, MA, USA}

\authorinfo{Further author information: (Send correspondence to P.P.P.)\\P.P.P.: E-mail: ppedersen@phys.ethz.ch\\}

\begin{document} 
\maketitle

\begin{abstract}
    We present the photometric performance of SPIRIT, a ground-based near-infrared InGaAs CMOS-based instrument (1280 by 1024 pixels, 12~\unit{\um} pitch), using on-sky results from the SPECULOOS-Southern Observatory during 2022 -- 2023. SPIRIT was specifically designed to optimise time-series photometric precision for observing late M and L type stars. To achieve this, a custom wide-pass filter (0.81 -- 1.33~\unit{\um}, \textit{zYJ}) was used, which was also designed to minimise the effects of atmospheric precipitable water vapour (PWV) variability on differential photometry. Additionally, SPIRIT was designed to be maintenance-free by eliminating the need for liquid nitrogen for cooling.
    We compared SPIRIT's performance with a deeply-depleted (2048 by 2048 pixels, 13.5~\unit{\um} pitch) CCD-based instrument (using an \textit{I+z'} filter, 0.7 -- 1.1~\unit{\um}) through simultaneous observations. For L type stars and cooler, SPIRIT exhibited better photometric noise performance compared to the CCD-based instrument. The custom filter also significantly minimised red noise in the observed light curves typically introduced by atmospheric PWV variability. In SPIRIT observations, the detector's read noise was the dominant limitation, although in some cases, we were limited by the lack of comparison stars.
\end{abstract}


\section{INTRODUCTION}
\label{sec:intro}  

The infrared domain is broadly defined between the wavelengths of 750~nm to 0.3~mm. This domain can reveal new spatial and temporal information invisible to other wavelengths. However, instruments in this domain are traditionally very costly and can come with a number of disadvantages due to inherent detector characteristics and environmental contributions such as thermal background and atmospheric absorption/radiation.

Most astronomical cameras today operate solely in the visible, predominantly using the silicon based Charged-Coupled Device (CCD) as the underlying technology.\cite{boyle1970charge} Over many decades, they have been well optimised to present low detector noise characteristics as well as high quantum efficiency (QE) and linearity. An alternative detector type, the Complementary Metal-Oxide-Semiconductor (CMOS), has recently presented similar noise and photometric qualities as CCDs, in addition to significantly faster readouts.\cite{alarcon2023scientific}

However, with CCD and CMOS detector's photosensitive material still based on silicon, its 1.1~eV band gap prevent photon absorption beyond 1100~nm. To push further into the infrared, one must consider materials that go below silicon's band gap, such as HgCdTe, and InGaAs.\cite{beletic2008teledyne, sullivan2013precision, sullivan2014near, simcoe2019background}  To produce an infrared detector, these materials are often hybridised onto existing silicon based CMOS readout circuitry designs. Larger demand for infrared cameras in industrial and military applications have primarily dictated the development direction of such cameras.

In the near-infrared (750~nm to 2.5~\unit{\um}), the majority of ground and space-based astronomy facilities have opted for the Hawaii HxRG line of detectors by Teledyne.\cite{beletic2008teledyne} They present state-of-the-art characteristics but with high financial, engineering, and maintenance cost when compared to the visible domain. Recently, a handful of InP substrate removed InGaAs off-the-shelf detectors and cameras have appeared on the market with competing specifications and increased appropriateness for ground-based robotic astronomy facilities due to non-liquid nitrogen cooling requirements.

With these recent advances, the SPECULOOS\cite{gillon2018searching, Delrez2018b, murray2020photometry, Sebastian20, 10.1117/12.2563563, burdanov2022speculoos} (Search for habitable Planets EClipsing ULtra-cOOl Stars) survey began an investigation into an infrared instrument to advance the survey's goals.\cite{pedersen2023optimised} Specifically, we sought a camera and filter combination which would optimise the photometric precision of observing ultra-cool dwarfs, whilst also minimising the unwanted effects of atmospheric precipitable water vapour (PWV) variability on the final data products -- time-series light curves.\cite{pedersen2022precise}

In this text, we detail the modelling, characterisation, and on-sky results of an infrared filter and camera combination called SPIRIT (SPeculoos InfraRed Imager for Transits) at SPECULOOS-Southern Observatory (SSO) during 2022 -- 2023. Both modelling and on-sky observations were done with reference to the existing deeply-depleted Si CCD (Andor iKon-L 936 BEX2-DD) based instrumentation at SSO.

\section{NEAR-INFRARED FEASIBILITY}\label{sec:nir-fes}

\begin{figure}[t!]
  \begin{center}
  \includegraphics[width=0.95\textwidth]{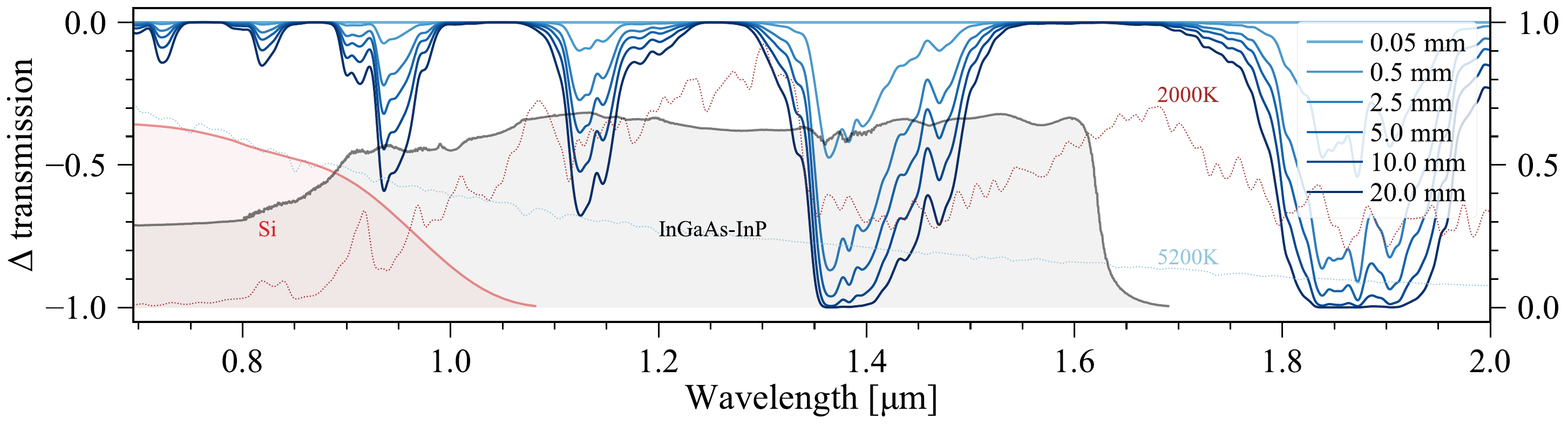}
  \end{center}
  \caption{The fractional change of atmospheric transmission (left-hand axis) from the visible to near-infrared, at airmass 1, from a PWV of 0.05~mm to 20~mm is shown via a series of low resolution atmospheric spectra.\cite{noll2012atmospheric, jones2013advanced} 5200~K and 2000~K stellar spectra are superimposed as dotted lines, from PHOENIX BT-Settl models.\cite{lim2015pysynphot, 2012RSPTA.370.2765A} Two total system efficiencies (right-hand axis) of a deeply-depleted Si based CCD and an InGaAs-InP based CMOS camera on a SSO Ritchey-Chrétien telescope are also shown.}
  \label{fig:h2ospectra}
\end{figure}

From the ground, atmospheric absorption due to PWV plays a significant role in the underlying accuracy of time-series light curves in the near-infrared. Fig.~\ref{fig:h2ospectra} shows the wavelengths PWV absorption is present along the total system efficiencies of the near-infrared instrumentation in consideration here -- a deeply-depleted Si based CCD and an InGaAs-InP based CMOS camera on a Ritchey-Chrétien telescope (as in SSO).

To assess the photometric feasibility of transitioning to the infrared, we first evaluated the bandpasses which would minimise the adverse effects of PWV variability. We then modelled the photometric precision of the relevant bandpasses for observing ultra-cool dwarfs from SSO.

\subsection{Mitigating precipitable water vapour effects: custom wide-pass filter}

To avoid the effects of PWV variability, the bandpass of choice must display a matching change of stellar flux, as a function of atmospheric PWV, for all the stars used during the differential photometry process. For the \textit{I+z'} bandpass, in use with existing deeply-depleted Si CCD based instrumentation at SSO, the PWV effect is exacerbated when there's a significant effective temperature difference between the target star and comparison stars used to produce the final light curve.

Ultra-cool dwarfs are often the coolest stars in a SPECULOOS field of view, with the majority of comparison stars (with a median effective temperature around 5200~K) significantly warmer.  However, despite this significant difference, there are bandpasses which can enable a equivalent change of stellar flux for the effective temperatures often used as part of the differential photometry process at SPECULOOS.\cite{10.1117/12.2561467, pedersen2022precise}

\begin{figure}[t!]
  \begin{center}
  \includegraphics[width=0.75\textwidth]{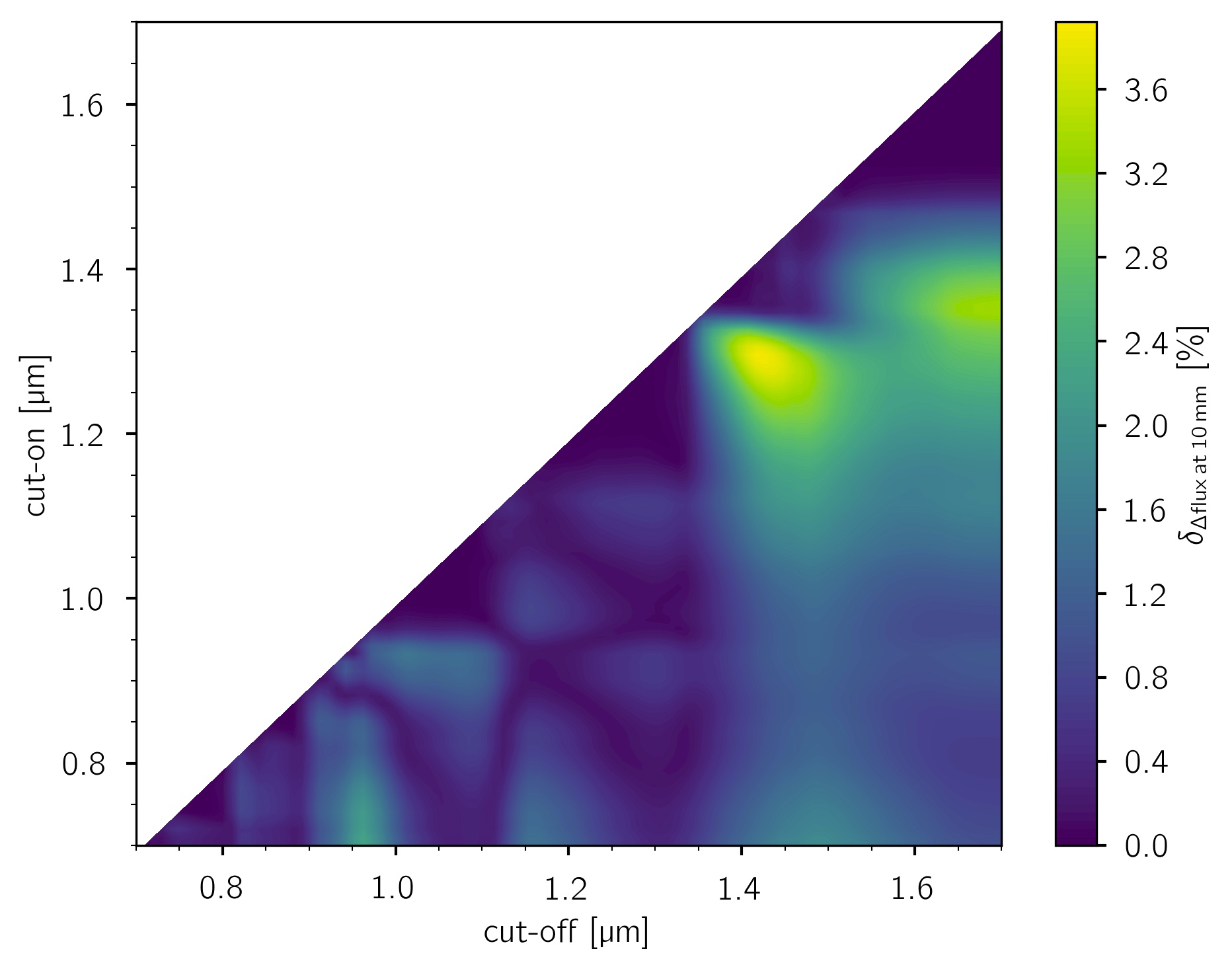}
  \end{center}
  \caption{The range of cut-ons and cut-offs for a bandpass observed with an InGaAs-InP based CMOS camera (see Fig.~\ref{fig:h2ospectra}), with the colour map representing regions where PWV variability poses issues to the differential photometry process, quantified by $\delta_\mathrm{\Delta flux~at~10~mm}$.}
  \label{fig:cut-on-off-delta-flux}
\end{figure}

To quantify the strength of the PWV effect, we developed a metric that models the change in stellar flux across a range of atmospheric PWV values (0.5~--~10~mm) and effective temperatures (2000~--~6000~K). This metric, denoted by $\delta_\mathrm{\Delta flux~at~10~mm}$, represents the standard deviation of the percentage change in stellar fluxes at 10~mm relative to 0.5~mm, calculated for all possible bandpasses in the near-infrared range detectable by an InGaAs-InP based CMOS camera. The results are illustrated in Fig.~\ref{fig:cut-on-off-delta-flux}.

From this, we identified local minima at and below 1.33~\unit{\um}, detailed in Tab.~\ref{tab:delta-flux-minimas}.  For SSO, photometry beyond the water absorption feature at 1.4~\unit{\um} was not deemed suitable for SPECULOOS' survey mode. The higher sky background flux and the efficiency cut-off of an InGaAs-InP detector would not yield better photometry than the bandpasses identified below 1.33~\unit{\um}, and thus was not considered here.

\begin{table}[t!]
  \caption{Local minimas identified in Fig.~\ref{fig:cut-on-off-delta-flux}. The metric for the \textit{I+z'} bandpass observed with deeply-depleted Si based CCD is also displayed.} 
  \label{tab:delta-flux-minimas}
  \begin{center}       
    \begin{tabular}{lrrr}
      \hline
      Name & Cut-on [\unit{\um}] & Cut-off [\unit{\um}] & $\delta_\mathrm{\Delta flux~at~10~mm}$ [\%]\\ 
      \hline
      \textit{J$_{MKO}$} & 1.17 & 1.33 & 0.06 \\
      \textit{YJ} & 0.96 & 1.33 & 0.09 \\
      \textit{zYJ} & 0.81 & 1.33 & 0.11 \\
      \textit{I+z'} (CCD) & 0.70 & 1.10 & 1.29 \\
      \hline
      \end{tabular}
  \end{center}
\end{table}

We found three bandpasses of interest, \textit{J$_{MKO}$}\cite{simons2002mauna}, \textit{YJ}, and \textit{zYJ}. The first bandpass is commonly found amongst near-infrared instrumentation, the remaining two bandpasses have not been found in literature as a singular bandpass. In the interest of maximising photometric precision for SPECULOOS targets, we evaluated the bandpasses with the methods detailed in the next section, and found the widest bandpass, \textit{zYJ}, presented a photometric advantage over \textit{YJ} and \textit{J$_{MKO}$} for the SPECULOOS target list\cite{Sebastian20}, despite the increased sky background in \textit{zYJ}.

\begin{figure}[t!]
  \begin{center}
  \includegraphics[width=0.75\textwidth]{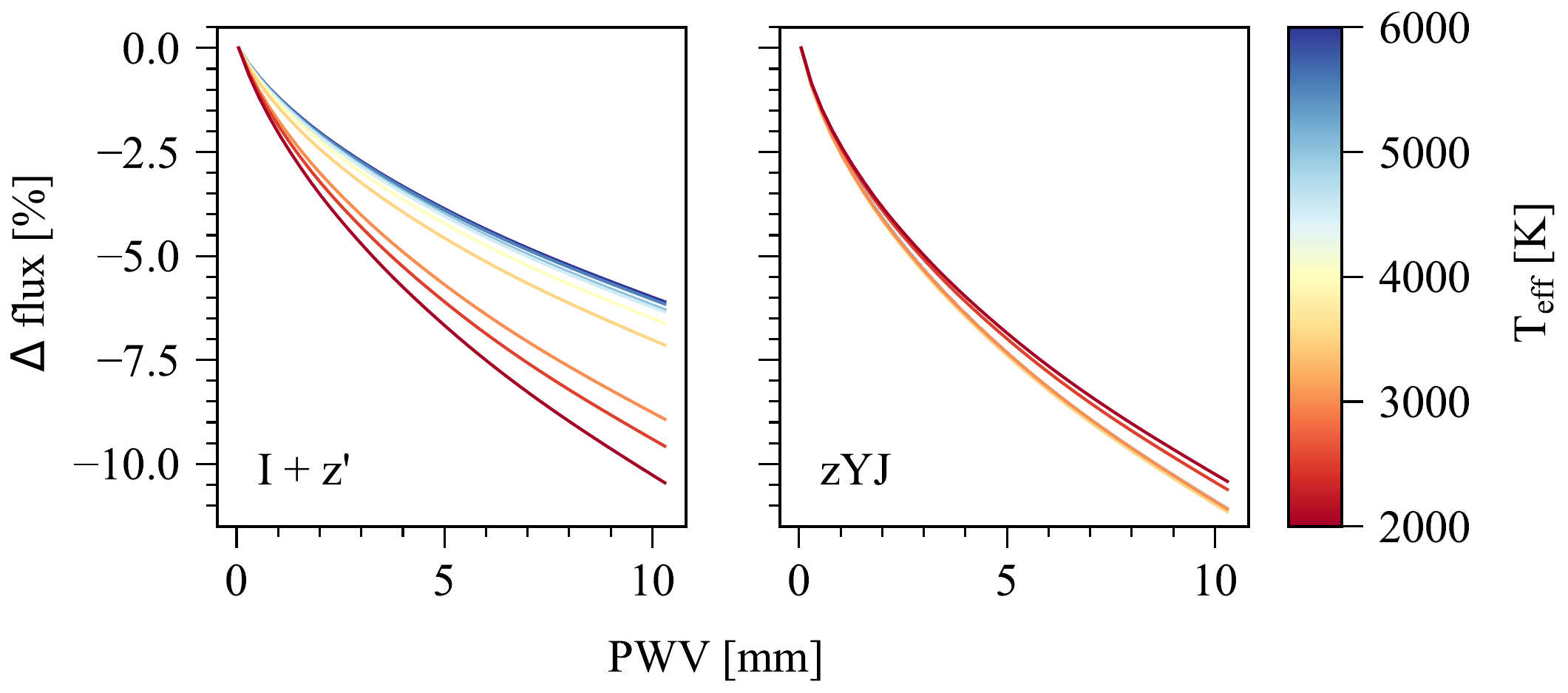}
  \end{center}
  \caption{The change in flux, $\mathrm{\Delta flux}$, as a function of PWV for different temperature stars (from 6000~K to 2000~K in steps of 500~K) as modelled through two observational bandpasses, \textit{I+z'} and \textit{zYJ}, with respect to a 0.05~mm PWV atmosphere at airmass 1.}
  \label{fig:pwv_effect_I+z_zYJ}
\end{figure}

Fig.~\ref{fig:pwv_effect_I+z_zYJ} shows the change of stellar flux, as a function of atmospheric PWV, for the \textit{I+z'} bandpass as observed with a deeply-depleted Si CCD based camera, and \textit{zYJ} observed with an InGaAs-InP based CMOS camera. The significant reduction in the spread of $\mathrm{\Delta flux}$ should lead to a reduced effect of false short and long variability in the final light curves induced by PWV variability.

\subsection{Modelling photometric precision: \textsf{mphot} python package}\label{sec:mphot}

Following an exposure from a ground-based facility, each pixel of a detector accumulates a total count of photoelectrons. These electrons originate from the target ($N_\star$), sky/environmental sources ($N_{sky} + N_{thermal}$), and inherent dark current in the detector ($N_{dark}$). All electron counts follow Poisson statistics and increase linearly over time.

Additionally, scintillation noise due to the Earth's turbulent atmosphere, $\sigma_{scint}$, is also present and a function of target brightness, airmass, and telescope aperture. We adopted the form of $\sigma_{scint}$ from Ref.~\citenum{osborn2015atmospheric}. During a pixel's readout, a time-independent noise is also added, $N_{read}$. The precision of observing a target thus follows

\begin{equation}\label{eq:pre}
    \sigma_{precision} = \frac{\sqrt{N_{\star} + \sigma_{scint}^2 + n_{pixel} \times
    (N_{sky} + N_{thermal} + N_{dark} + N_{read}^2)}} {N_{\star}},
\end{equation} 

where $n_{pixel}$ is the number of pixels that encircles the point spread function (PSF) of the target. For aperture photometry, the aperture radius is normally 2-3 times the full-width-half-maximum (FWHM) of the PSF.\cite{Nutzman2008, murray2020photometry}

To calculate the values of $N_\star$ and $N_{sky}$, we employed model stellar spectra, transmission models, and radiative models of our atmosphere. We developed a tool called \textsf{mphot}\footnote{\url{https://github.com/ppp-one/mphot/}} to integrate these models over the system response of the instrumentation and telescope, enabling us to generate interpolatable grids of stellar fluxes and sky radiances as a function of atmospheric PWV and airmass. This allowed us to model the precision for the entire SPECULOOS target list.\cite{Sebastian20}

The effective temperature range of SPECULOOS' target list spans from 1278~K to 3042~K. To model their spectra, we used PHOENIX BT-Settl stellar models for temperatures down to 2000~K and ATMO 2020 chemical equilibrium models for temperatures below 2000~K.\cite{2012RSPTA.370.2765A, laidler2008pysynphot, phillips2020new} To build a stellar spectrum, we used a representative set of stellar parameters from Ref.~\citenum{pecaut2013intrinsic}\footnote{Specifically, the updated values from \url{https://www.pas.rochester.edu/~emamajek/EEM_dwarf_UBVIJHK_colors_Teff.txt}, version 2021.03.02}, assigning a metallicity index of 0 to each spectrum. Synthetic brightnesses were modelled and calibrated using J magnitudes from the 2MASS survey.\cite{2MASS}

We obtained 273 atmospheres from ESO's 2400~m SkyCalc Sky Model\cite{noll2012atmospheric, jones2013advanced} through a permutation of airmasses between 1 and 3 with 0.1 increments and PWV values between 0.05~mm and 30~mm (specifically, 0.05, 0.1, 0.25, 0.5, 1.0, 1.5, 2.5, 3.5, 5.0, 7.5, 10.0, 20.0, and 30.0~mm). We subsequently generated 107 model stellar spectra between 450 K and 36,500 K, producing grids with a total of 29,211 values to interpolate between.

To obtain a value from Eq.~\ref{eq:pre}, we must also know properties inherent to the detector/optics/site, including $N_{dark}$, $N_{read}$, pixel well depth, pixel pitch/plate scale, $N_{thermal}$, seeing conditions, and exposure time. To estimate the exposure time, we modelled the target star's brightest pixel to fill 70\% of the well depth, assuming a 2D Gaussian distribution based on the seeing to model the flux distribution on the detector.

For the detector properties, following an extensive market search, we identified a suitable commercial-off-the-shelf InGaAs-InP based CMOS camera for the SSO: the Princeton Infrared Technologies (PIRT) 1280SciCam.\footnote{Marketed specifications at time of interest (2018-02): \url{https://web.archive.org/web/20180227174252/http://www.princetonirtech.com/products/1280scicam}} Our photometric precision modelling indicated that, without modification to the optical configuration at SSO, the 1280SciCam's performance with the \textit{zYJ} bandpass would outperform the existing Si CCD-based instrument with the \textit{I+z'} bandpass for targets with temperatures cooler than 2550~K, under median conditions at Paranal, Chile. Its characterised specifications are detailed in the next section.

\section{REAL LIFE PERFORMANCE}

We acquired a 1280SciCam in late-August 2021, after 1.5 years of pandemic procurement related delays.  Between September and November 2021, we characterised basic detector parameters and produced an ASCOM camera driver in preparation for use at SSO.  From March 2022, the camera and the \textit{zYJ} bandpass were on-sky at SSO until the camera suffered a vacuum seal leak in late 2022, and was removed from operation in early 2023.

\subsection{Characterisation}

By default, the 1280SciCam produced 1280$\times$1024 14 bit images at a frequency of 1/(frame time). The frame time was user-settable and defined as the time from the start of an exposure to frame read.  Leveraging its Capacitive Transimpedance Amplifier (CTIA) based CMOS architecture, with Analog to Digital Converters (ADCs) every 8 rows, the camera enabled fast global readout without the need for a mechanical shutter.  Communication with the 1280SciCam was performed via the Camera Link protocol with a Bitﬂow Axion 1xE frame grabber and a Camera Link to fiber extender (EDT VisionLink RCX).

While the camera could be programmed to trigger single exposures on an external signal, this approach was found to increase read noise. We thus employed the continuous mode with a dynamic frame time set by the ASCOM driver, where the frame time was set to be the exposure time plus a 100~ms buffer. We did not optimise the buffer time duration, as it was negligible compared to our typical exposure times (36.5~s median exposure time for the modelled SPECULOOS target list). However, we did observe that shorter buffer times occasionally introduced image banding artefacts, further investigation is needed to fully understand this and its effect on photometric quality.

\begin{table}[!th]
  \centering
  \caption{PIRT 1280SciCam characterised specifications for the default SOC operating at $-$60~\si{\celsius}.}
  \begin{tabular}{lrrr}
      \hline
      Parameter & Unit & Values & Notes\\
      \hline
      Array format & pixel & 1024~$\times$~1280 & \\
      Detector material &  & InGaAs & \\
      Detector substrate &  & InP - removed & \\
      QE (across \textit{zYJ}) & \% & 77 & App.~\ref{app:qe} \\
      Pixel pitch & \unit{\um} & 12 & \\
      Pixel fill factor & \% & $>$ 99 & \\
      Readout duration & s & $\sim$ 0.1 & Set by driver\\
      Readout method & & CDS &  \\
      Data output & bits & 14 & \\
      Gain & e$^-$/ADU & 5.092 $\pm$ 0.004 & Sec.~\ref{sec:ptc}\\
      Well capacity & e$^-$/pixel & $\sim$ 56470 & Sec.~\ref{sec:ptc}\\
      Read noise & e$^-$/pixel & 89.95$\pm$0.15 & Sec.~\ref{sec:ptc}\\
      Dark current & e$^-$/s/pixel & 57.2$\pm$0.8 & Sec.~\ref{sec:dc}\\
      Thermal flux & e$^-$/s/pixel & $<$ 60 & Sec.~\ref{sec:dc}\\
      Bad pixels & \% & 0.325 & Sec.~\ref{sec:bad-pixels}\\
      Non-linearity & \% & $<$ 0.95 & App.~\ref{app:lin} \\
      Persistence & \% & $<$ 1.1 & App.~\ref{app:per} \\
      Field of view & ' & 5.3~$\times$~6.6 & SSO telescope\\
      \hline
  \end{tabular} 
  \label{tab:final-specs}
\end{table}

On our unit, the manufacturer provided three ``Specific Operating Conditions" (SOCs) which differed in bias voltage and undisclosed use of the readout electronics.  We settled on using the default SOC at 3.1~V bias voltage, as the other experimental modes revealed issues or underperformed with respect to the default mode. 

For the remainder of this text, we describe the performance of the 1280SciCam operating at $-$60~\si{\celsius} using its 4-stage peltier (hot-end liquid cooled at 15~\si{\celsius}) with the default SOC. The final characterised values are shown in Tab.~\ref{tab:final-specs}. In lab, we characterised the gain, read noise, well depth, and bad pixels. At SSO, we characterised the dark current and thermal flux. The QE, non-linearity, and persistence were measured by PIRT (see App.~\ref{app:qe}--\ref{app:per}). The \textit{zYJ} filter transmission (see App.~\ref{app:filter}) was measured by its manufacturer, Brinell Vision.

\subsubsection{Photon transfer curve -- bulk gain, read noise, and well depth}\label{sec:ptc}

\begin{figure}[t!]
  \centerline{\includegraphics[width=1\textwidth]{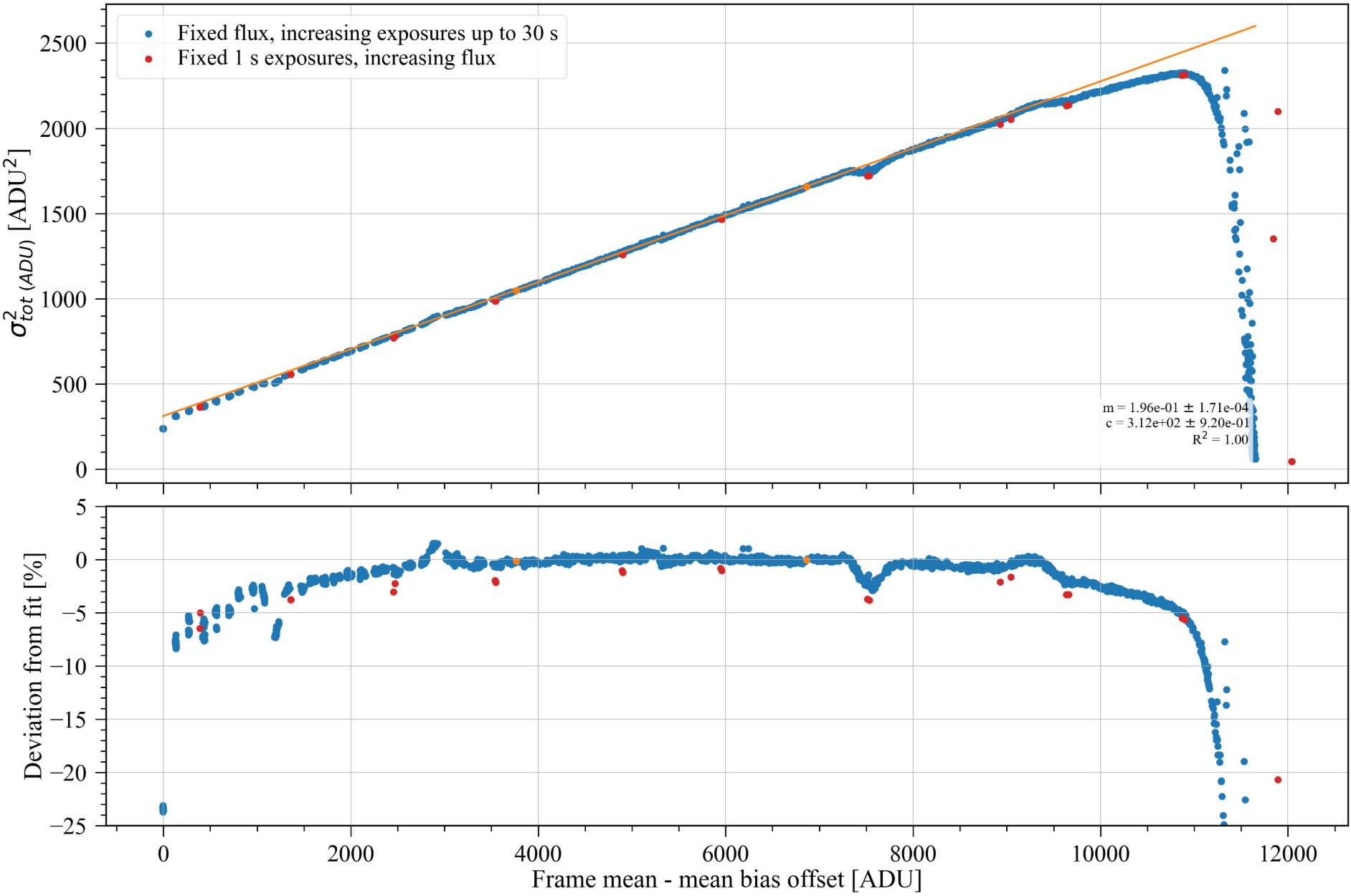}}
  \caption{Mean-variance photon transfer curve, in blue for 2400 pairs of flat fields (after removing bad pixels) of increasing exposure up to 30~s of a constant flux source, and in red for 24 pairs of flat fields of 1~s fixed exposure period and increasing flux. Top: Variance of flat field pair difference (divided by 2) against flat field pair mean minus mean bias offset, with the line of best fit along a linear region of blue data points (denoted between the orange points). Bottom: Deviation from line of best fit for both datasets.
  }\label{fig:PTC}
\end{figure}

We employed the mean-variance photon transfer curve (PTC) method to determine the camera's bulk gain, read noise, and well depth. This approach involved measuring the signal variance in the difference between consecutive frame pairs, $2\sigma_{tot}^2$, as a function of changing flux.\cite{birch2022ingaas} We took the difference because it removed any fixed-pattern present in the frames. In total, we acquired 2400 images in front of an integrating sphere with a stabilised halogen lamp, with a 100 distinct exposure times ranging from 0 to 30~s. This yielded 24 frame pairs per exposure period. The resulting plot is depicted in Fig.~\ref{fig:PTC}.

Following Poisson statistics, the gain will be the inverse gradient of frame means against their variances, with the intercept equivalent to the detector's read noise. However, if the detector is non-linear, this method can lead to inaccurate estimates of these parameters.\cite{pain2003accurate} Furthermore, the read noise may exhibit a dependence on the signal, which would also result in incorrect parameter estimations using this approach.\cite{9144291}

Nonetheless, we were able to take a linear fit along the central region of the curve. Deviations from linearity were seen at low flux and high flux levels before saturation. Five distinct features were also observed deviating from the line of best fit, one near the center of the usable ADU range, followed by two pairs with opposing deviations. The source of these distinct deviations is unknown but thought to have a signal dependence rather than an exposure period dependence. Despite these deviations, they were not visible in linearity measurements (see App.~\ref{app:lin}).

Repeating the mean-variance PTC with a smaller set of exposures, we kept a fixed exposure period of 1~s and varied the flux output from the integrating sphere. By chance, we matched the mean flux level at the center of one of the distinct features, where we could see a similar deviation. We also observed that the overall variance was marginally lower compared to the previously determined line of best fit. The 1~s exposures still maintained a similar gradient, suggesting that the gain was similar along the central flux region, but with slightly lower read noise. Another notable difference was the region of saturation, which was $\sim$500 ADU larger than the previous dataset - suggesting that the well depth may be smaller for larger exposure periods.

Despite these deviations, we calculated the gain along the linear region to be 5.092$\pm$0.004~e$^-$/ADU, with a read noise of 89.950$\pm$0.154~e$^-$. Using the same gain, we found the read noise calculated through a series of bias frames (100 $\times$ 100~\unit{\us} exposures) was lower, with a median read noise of 78.5~e$^-$, suggesting lower noise behaviour at shorter exposure times or well depths.

To quantify the full well, we considered the point on the PTC curve where the variance starts to decline (i.e. when the local gradient becomes negative), this gave us an indication of when the average pixel starts to decline in usability. This method yielded a saturation value of $\sim$11090~ADU, or 56470~e$^-$. For the 1~s mean-variance PTC, the 0 value variance point was $\sim$500 ADU higher, suggesting an additional 2500~e$^-$ for the average well depth. Although, with the gain potentially changing as a function of signal level, this electron value is likely to be different in reality.

\subsubsection{Dark current and thermal flux}\label{sec:dc}

The relative proportions of the elements that make up the 1280SciCam detector, indium gallium arsenide, dictate the cut-off and stability of the material, which in turn affects the dark current produced per pixel and the thermal flux it is sensitive to. PIRT have chosen a stable configuration of InGaAs (In$_{0.53}$Ga$_{0.47}$As) which minimises the strain in the material. In this configuration, the cut-off (50\% from maximum QE) wavelength was found to be around 1624~nm at $-$60~\si{\celsius}, as shown in Fig.~\ref{fig:h2ospectra}. 

\begin{figure}[ht]
  \centerline{\includegraphics[width=0.90\textwidth]{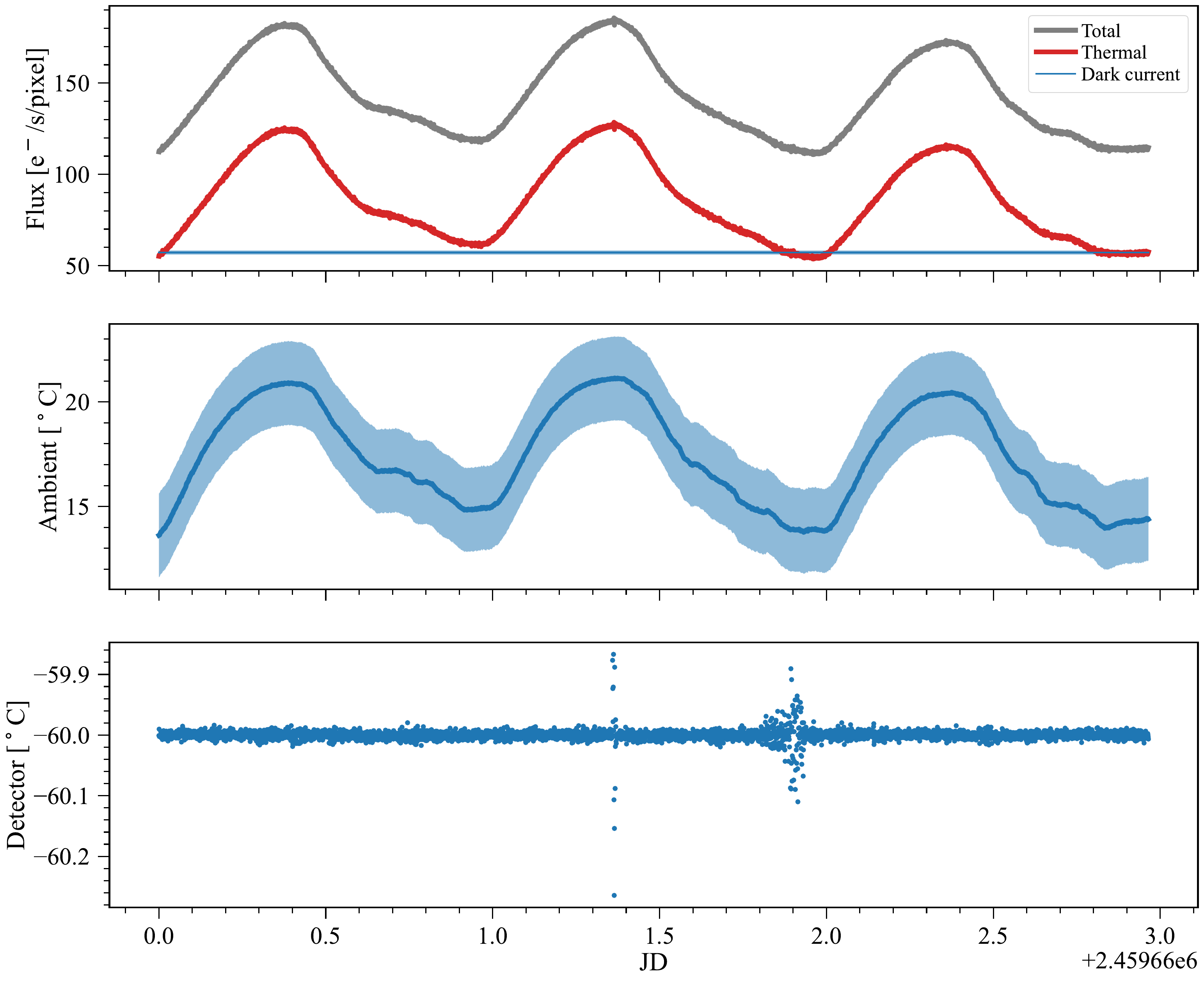}}
  %
  %
  \caption{Top: 60~s dark frames taken over 3 days (using the median of the frame). Split into the total flux (gray), thermal flux (red), and dark current (blue).  Middle: Ambient temperature from a sensor near the edge of the dome with an assigned error of $\pm$1.5~\si{\celsius}, offset by 45~minutes to account for the time for SPIRIT/telescope needed to thermally equilibrate. Bottom: Detector temperature, which was set to run at $-$60~\si{\celsius}.
  }\label{fig:bk-dc-time}
\end{figure}

To separate the dark current from thermal flux, we recorded a series of dark frames over a range of ambient temperatures, between approximately 14 and 21~\si{\celsius}. Assuming a blackbody emission from the environment in front of the detector, we could estimate the thermal flux using Planck's law integrated over the sensitive wavelengths of the 1280SciCam.  From this, the true dark current could be revealed, as shown in Fig.~\ref{fig:bk-dc-time}.

The dark current was estimated to be around 57.2$\pm$0.8~e$^-$/s/pixel, similar to the value found in Ref.~\citenum{birch2022ingaas}. During nightly operation at Paranal, the median estimated thermal flux was around 32~e$^-$/s/pixel derived using temperature values from our outdoor weather station, with the 1 and 99 percentile being 10 and 60~e$^-$/s/pixel, respectively.  For comparison, a HgCdTe 1.75~\unit{\um} cut-off detector, with the same 12~\unit{\um} pixel pitch, would be estimated to have around 500~e$^-$/s/pixel of thermal flux, and around 4000~e$^-$/s/pixel in dark current if the detector operated at the same $-$60~\si{\celsius} as the 1280SciCam under the same median ambient temperatures during operation of 14~\si{\celsius}. To reduce the thermal flux to sub-1~e$^-$/s/pixel received by SPIRIT, the optical train would have to be cooled to $-$19~\si{\celsius} from our estimates.

\subsubsection{Bad pixels}\label{sec:bad-pixels}

The cause for underperforming pixels can be due to any number of reasons, such as: poor pixel bump bond connectivity, vignetting, addressing faults in the mux, severe sensitivity deficiency, poor signal linearity, low full well, excessive noise and high dark current.\cite{smith2014bad} Rather than investigating the underlying causes, our approach focused on categorising pixels that deviated from the mean, $\mu$, by more than 3$\sigma$ under dark and uniform illumination conditions. This allowed us to characterise the distribution and approximate quantity of defective pixels on the detector.

A master dark frame was formed from 100 frames at 30~s exposures, taking the median of each pixel in the stack, with the frame's median filling 11\% of the usable well defined in the previous subsection. The master flat was similarly formed from 100 frames at 14~s exposures in front of an integrating sphere, frame's median reaching 50\% of the well, once again taking the median of each pixel in the stack. Each master frame was subtracted from a master bias frame (formed from 100 $\times$ 100~\unit{\us} exposures).

\begin{table}[t]
  \centering
  \caption{Percentage value of the frame's underperforming pixels ($\pm3\sigma$ outside of the frame's mean) for a master flat and master dark (master bias subtracted). Results divided between the assessed frames, and the proportions below and above the usability thresholds, and combined total of underperforming pixels from both frames.}
  
  \begin{tabular}{lrrr}
      \hline
      ~ & Lower [\%] & Upper [\%] & Total [\%]\\
      \hline
      Flat & 0.046 & 0.225 & 0.271\\
      Dark & 0.052 & 0.254 & 0.306\\
      Combined &  &  & 0.325\\
      \hline
  \end{tabular}
  
  \label{tab:bad-pixels}
\end{table}

Tab.~\ref{tab:bad-pixels} presents the proportion of underperforming pixels for each assessed frame, along with the total percentage of identified bad pixels. Notably, a greater number of hot pixels than cold pixels were found with this detector. However, defining an exact number of hot and cold pixels is challenging due to the overlapping distributions of the flat and dark frames, wherein some of the $\mu - 3\sigma$ values of the flat frame overlap with some of the $\mu + 3\sigma$ values of the dark frame. Similarly, the relative proportions would vary depending on the flux levels in the respective dark and flat frames given the flux level's influence on $\sigma$ and $\mu$.

An examination of a series of dark frames with increasing exposure times revealed that a large proportion of bad pixels were associated with a negative or zero dark current. Furthermore, during operation, we observed a small proportion of pixels ($\sim$300 pixels) exhibiting telegraphic noise, where the pixel values would shift by thousands of ADU from frame to frame.

The distribution of bad pixels was found to be relatively uniform across the detector, with no apparent clustering in specific regions. This distribution will likely pose issues for the differential photometry process, discussed in the next section. Although the total percentage of bad pixels found here (0.325\%) met our requirement of $<$0.5\%, we recommend that future procurements incorporate a metric that quantifies the distribution of bad pixels as part of the requested specification to minimise the risk of having bad pixels in the stars of interest for photometry.

\subsection{On-sky performance}

\begin{figure}[t!]
  \centerline{\includegraphics[width=0.8\textwidth]{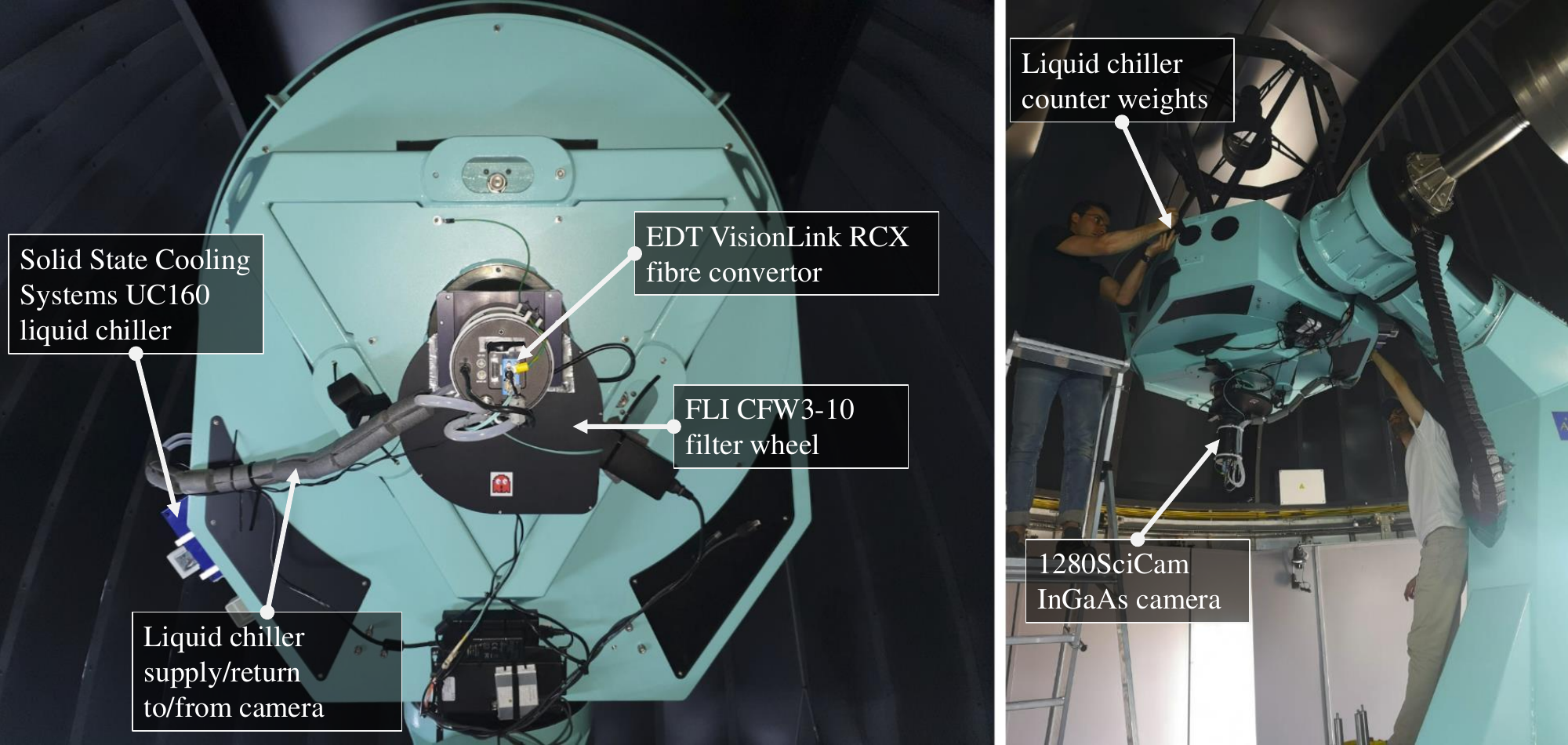}}
  %
  %
  \caption{Left: Labelled equipment forming SPIRIT, attached to Callisto. Right: Process of balancing the declination axis, to counteract the additional weight from the liquid chiller. Images of the modified liquid chiller are shown in App.~{\ref{app:chiller}}.
  }\label{fig:installed-system}
\end{figure}

SPIRIT was integrated at SSO in early 2022, where it replaced the Si CCD-based instrument on Callisto, one of the four 1~m class telescopes at SSO, as shown in Fig.~\ref{fig:installed-system}.  A neighbouring SSO telescope, Europa, installed with SSO's normal Si CCD-based instrumentation, was used for simultaneous observations to compare with SPIRIT.

\subsubsection{Light curves from a L5V type star}

We simultaneously observed a 1694$\pm$104K target from the SPECULOOS target list\cite{Sebastian20}, Sp1507-1627, with a J magnitude of 12.830$\pm$0.027.

\begin{figure}[t!]
  \centerline{\includegraphics[width=0.8\textwidth]{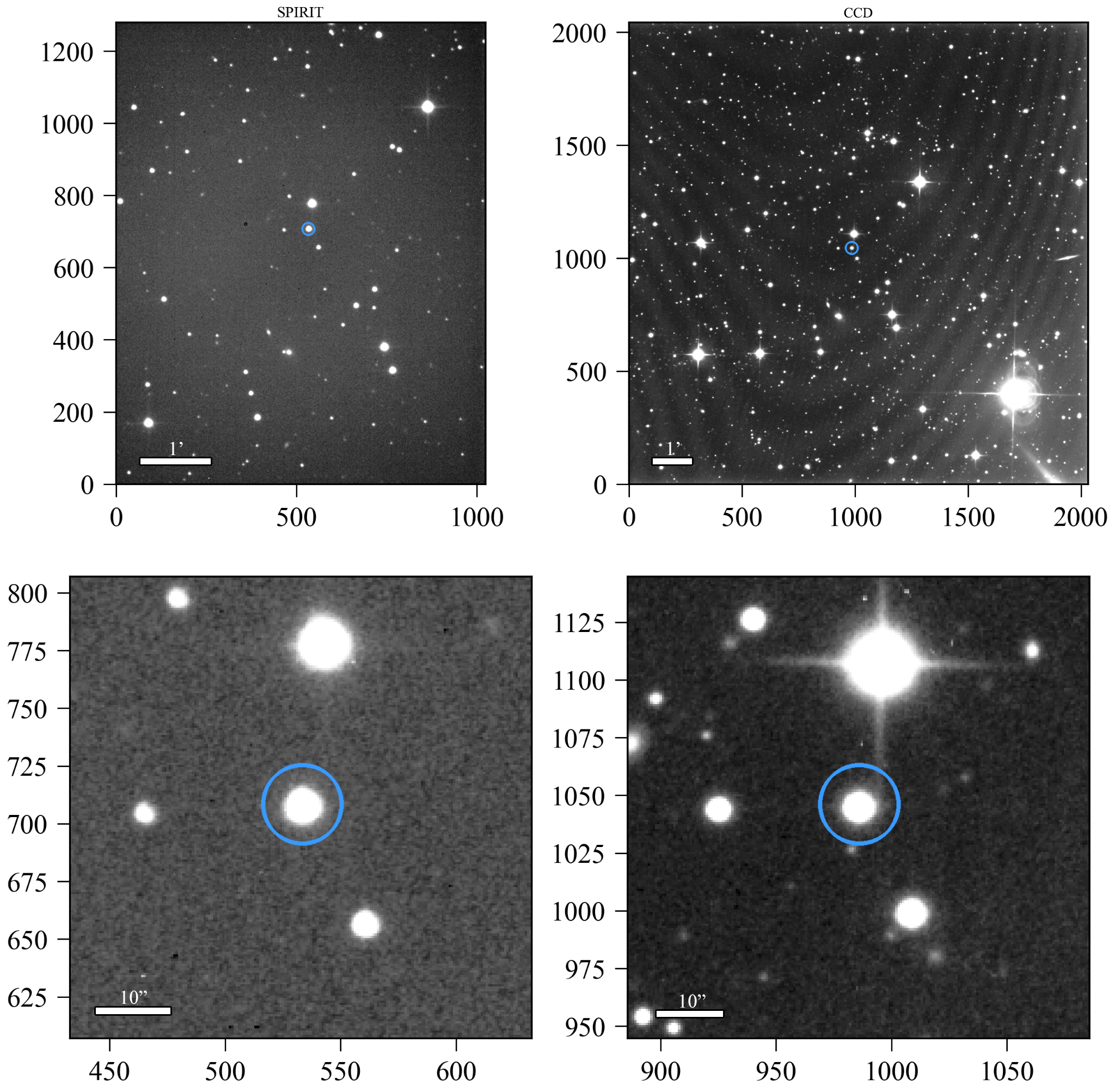}}
  %
  %
  \caption{
      Reduced stacked images of the Sp1507-1627 field of view with target circled in blue. Top-left: SPIRIT's full field of view. Bottom-left: SPIRIT's cropped in field of view, with the target star centred. Top-right: Europa's full field of view with the existing instrumentation. Bottom-right: Europa's cropped in field of view, with the target star centred.
  }\label{fig:stacks}
\end{figure}

A bad pixel map was computed for SPIRIT from the darks and flats, following the same methodology as described in Sec.~\ref{sec:bad-pixels}. We masked the bad pixels using an average of the surrounding pixels (excluding neighbouring bad pixels) around a bad pixel. We found it essential to use a master dark formed of matching exposure times to the science frames as part of the reduction process, as this reduced artefacts from bias structure and readout integrated circuit (ROIC) glow.

Upon inspection of the aligned stack, in Fig.~\ref{fig:stacks}, three key differences were noted. First, SPIRIT's reduced stack exhibited a lack of fringing, consistent with the findings of Ref.~\citenum{birch2022ingaas}. Second, a dark ring-like feature was observed in the stack, corresponding to a bright feature in the flats, located approximately 1' left of the target. This feature resulted in a region that was approximately 5\% dimmer than the surrounding area, although the cause of this remains unknown, it's likely deposition of additional material on the detector plane. Third, fewer stars were visible in the same angular area, which was expected given the higher noise floor of SPIRIT.

\begin{figure}[t!]
  \centerline{\includegraphics[width=0.9\textwidth]{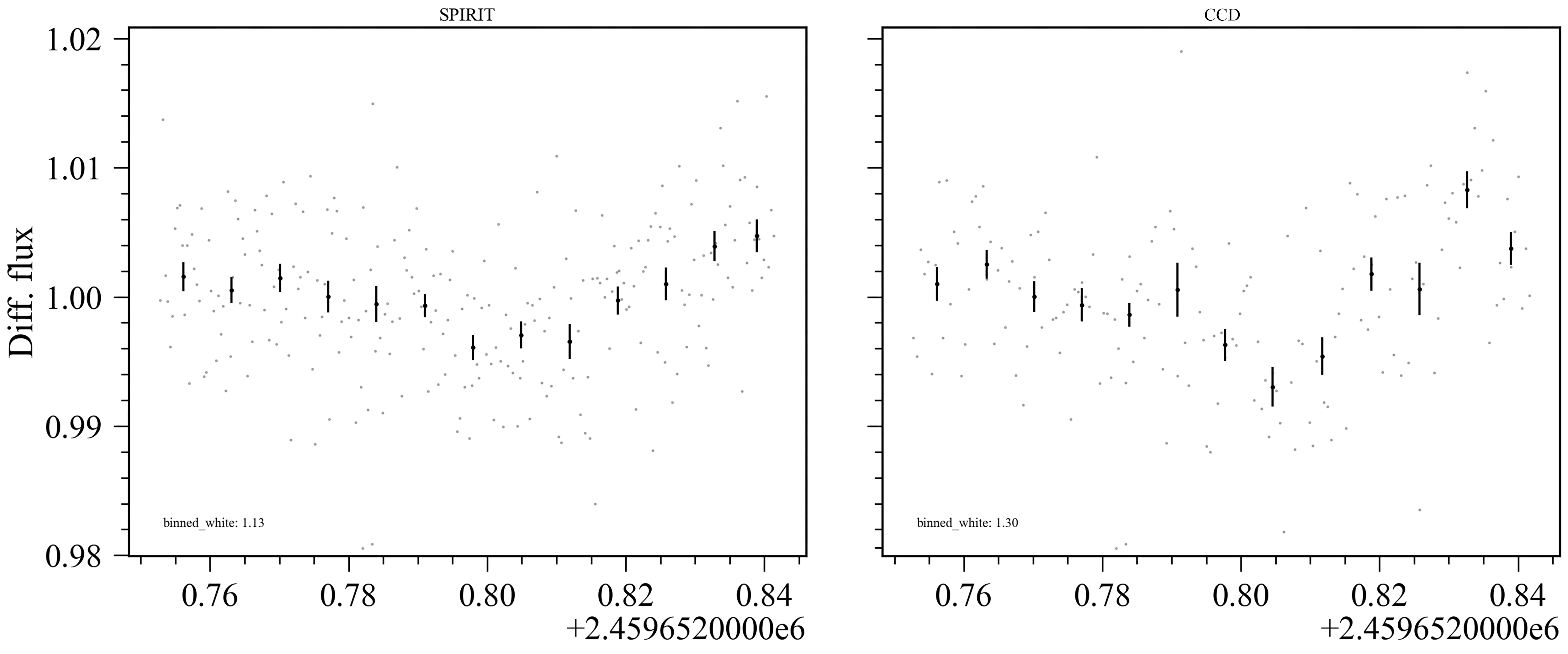}}
  %
  %
  \caption{
      10-minute binned light curves (in black, raw data in grey) of Sp1507-1627 taken on the night of 2022-03-13 under an aperture radius of 8.9 pixels for SPIRIT and 6.5 pixels for Europa. SPIRIT observed with the \textit{zYJ} filter (left) and the existing instrumentation on Europa with \textit{I+z'} (right).
  }\label{fig:cool-lc-only}
\end{figure}

The resulting light curves from the night's observation are shown in Fig.~\ref{fig:cool-lc-only}, which were processed using \textsf{Prose}\cite{garcia2022prose}. Using the algorithm described in Ref.~\citenum{broeg2005new}, 2 comparison stars were chosen for SPIRIT, and 7 for Europa (majority of which were outside SPIRIT's field of view). Choosing the same comparisons as SPIRIT for Europa's photometry yielded the same structure as shown but with slightly larger binned photometric noise.

To determine if the precision of the observation matched our models, we calculated the binned precision of all stars in the field which met the following criteria: stars whose peaks exceeded the background flux by at least three times, and did not saturate during the observation. We estimated their Vega magnitudes in the respective bandpasses using Vega's spectra from Astrolib PySynphot spectrum library.\cite{lim2015pysynphot}

SPIRIT's detector parameters were taken from Tab.~\ref{tab:final-specs} and from the darks taken the morning after the observations (dark current + thermal: 90~e$^-$/s/pixel). Europa's camera parameters were taken from its manufacturer and from measurements on site (gain: 1.072~e$^-$/ADU, read noise: 5.8~e$^-_{rms}$, dark current: 0.2~e$^-$/s/pixel). The median sky background for SPIRIT was measured to be 51~e$^-$/s/pixel, 20\% lower than the median models predicted. For Europa, the median sky background was 33\% lower than the models, suggesting that it was a better than average night for sky background. 

\begin{figure}[t!]
  \centerline{\includegraphics[width=0.95\textwidth]{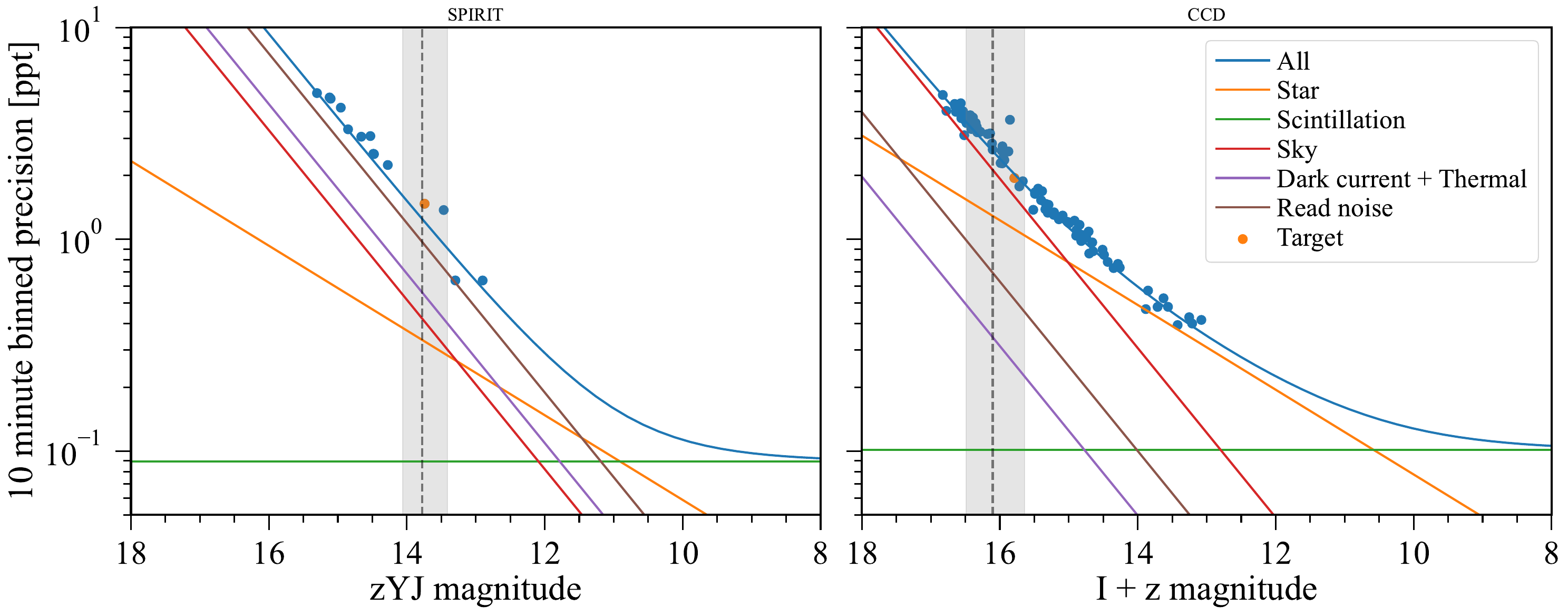}}
  %
  %
  \caption{
      Comparing the photometric 10-minute binned precision of SPIRIT and Europa with the observations made, with all the identified stars as blue points, and target star in orange, plotted against their respective magnitudes. Here, a larger set of aperture radii were used to encapsulate the majority of the target's flux, 12.16 pixels for SPIRIT, 11.31 for Europa. Vertical dashed lines show the expected magnitudes for a 1694~K star at 7.4~pc using the models utilised as part of the modelling in Sec.~\ref{sec:nir-fes}, the shaded grey region shows the expected magnitude range for $\pm$100~K from the target temperature. The coloured lines show the respective contributions of the noise model, with the blue line representing the total model.
  }\label{fig:noise-models}
\end{figure}

As illustrated in Fig.~\ref{fig:noise-models}, our results show a good agreement between the precision models and the observed data. We can see that Europa was limited by its sky background, whereas SPIRIT was limited by its read noise and potentially by its fewer number of comparison stars. This agreement showed SPIRIT surpass the photometric white noise of existing instrumentation. Modelling from Sec.~\ref{sec:mphot} suggest SPIRIT is likely to deliver improved photometric performance for SPECULOOS targets with temperatures below 2550~K. However, it is essential to acknowledge the uncertainty regarding the impact of red noise and bad pixels on photometric accuracy. In this particular observation, neither the target nor the comparison stars were affected by bad pixels.

We achieved a better median seeing with Europa, measuring 1.17", compared to 1.35" with SPIRIT. Initially, we expected SPIRIT to demonstrate marginally better seeing than the existing instrumentation, given its operating bandpass further in the infrared.\cite{Boyd:78} However, our observations revealed that SPIRIT consistently exhibited higher seeing. We now suspect this was due to minor mirror misalignment. 

Prior to this observation, we guessed an initial seeing of 1" for SPIRIT to estimate its exposure time. The exposure times for each instrument were: 30~s for SPIRIT and 36~s for Europa. SPIRIT's exposure time was calculated to fill 70\% of the well, based on our initial assumption of 1" seeing. However, for Europa, to avoid exceeding the 120~s cadence limit of SPECULOOS, Europa's exposure time was adjusted to achieve a lower well fill, to minimise the effect of red noise\cite{pont2006effect} and maintaining acceptable precision in binned data, made possible by Europa's low read noise.

Optimised exposure times for the night's observations could have yielded greater precision. However, neither telescope reached the 70\% well depth target. Even if Europa's exposure time had approached the 120~s cadence limit, it would not have reached the 70\% well depth. Nonetheless, if Europa's exposure time had matched the 120~s limit, the 10-minute binned photometric precision would likely have improved by an estimated 10\%. Similarly, if SPIRIT had reached 70\% of its available well depth (by doubling its exposure time), we would have seen an estimated 15\% improvement in its 10-minute binned precision. 

Furthermore, SPIRIT utilised more pixels in its aperture than Europa given its larger perceived seeing and smaller pixel plate scale, increasing the total contribution of read noise and dark current in its photometry. Optimising the plate scale of SPIRIT is recommended for future work -- in part to optimise the noise contribution per aperture, but also to optimise the number of comparison stars in its field of view. However, an investigation into the effect of flat-fielding/tracking errors is necessary before re-imaging the image plane.

There was a dip seen in both light curves, which is not believed to be a transit. If it were to be, Sp1507-1627 would be the coolest exoplanet host star after Trappist-1\cite{Gillon2017} found via the transit method. The dip was less pronounced in SPIRIT's light curve suggesting a lower sensitivity to the effect.  If a real feature, it was likely sourced from a stellar spot, whose contrast is lower in the infrared.\cite{pedersen2023optimised, almenara2022toi} We have successfully observed known transits with SPIRIT, such as the one detailed in Ref.~\citenum{10.1093/mnrasl/slad097}.

\subsubsection{PWV insensitivity}\label{s:pwv-insensitivity-obs}

\begin{figure}
    \centerline{\includegraphics[width=1\textwidth]{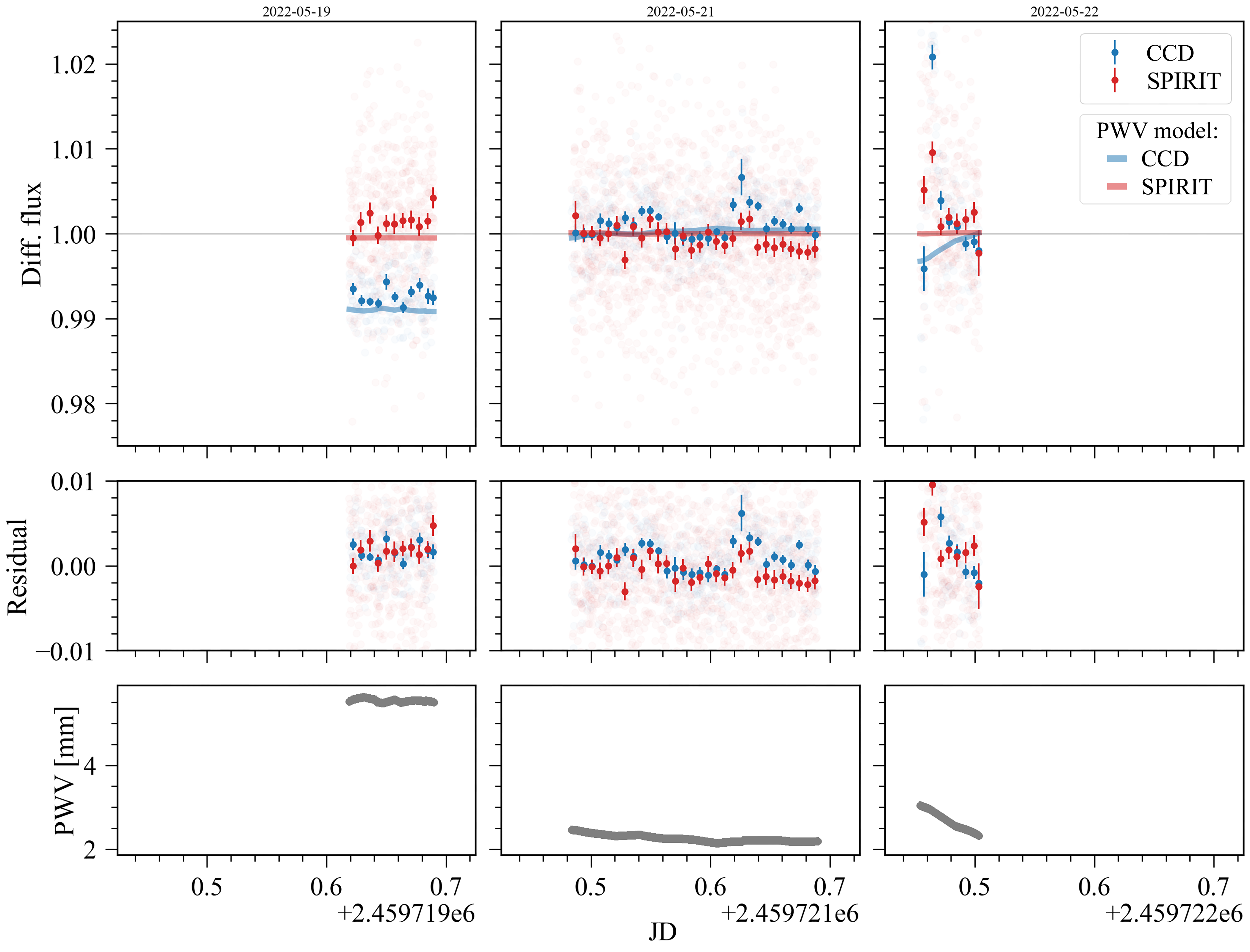}}
    %
    %
    \caption{
        Globally normalised light curves of a 2720~K target observed by SPIRIT (\textit{zYJ}) and Europa (\textit{I+z'}). Top row: 10-minute binned differential light curves as solid points, and raw data as fainter points, without PWV correction. SPIRIT's observations in red, and Europa's CCD in blue. Solid lines showing the expected trend induced by the PWV variability assuming a 5000~K comparison star, in the same respective colours as the observations. Middle row: The residual light curve after subtracting the expected trend induced by the PWV variability. Bottom row: PWV measurements made by the LHATPRO\cite{kerber2012water} at zenith.  Each x-axis has the same length to permit comparisons of feature changes over the same timescale. Missing data were due to either bad weather or technical downtime.
    }\label{fig:pwv-obs}
\end{figure}

The existing instrumentation with the \textit{I+z'} filter is highly sensitive to PWV changes during an observation.\cite{pedersen2022precise} To test the effectiveness of the produced \textit{zYJ} filter, we observed a known quiet target previously observed by SPECULOOS over a series of nights until a large PWV change occurred. The results of the observations are shown in Fig.~\ref{fig:pwv-obs}.

We observed a 2720$\pm$104~K target with a J magnitude of 11.840$\pm$0.022. It was observed by both SPIRIT and Europa, at an exposure time of 12~s and 26~s respectively. The available comparison stars for both fields were all around 5000~K. During the observation run, we observed a large PWV change, 5.6~mm to 2.1~mm, a $\Delta$PWV of 3.5~mm, which induced a 0.98\% change in Europa's light curve, and a 0.07\% change in SPIRIT's as suggested by the modelled change. The induced changes in Europa's light curve was a factor of 15 higher than that seen by SPIRIT. 

The residual light curves show agreement with one another, with a flare seen on the final night. This agreement is indicative that the PWV suppressing nature of the \textit{zYJ} filter design is working as expected for this example target. However, to have greater confidence in its performance, further observations with cooler targets would be preferred.

This result has implications for the future of ground-based infrared observations which are time-series orientated. If maximising photometric precision is the main objective of a survey, for example when searching for transits, then this work demonstrates that wider bandpasses beyond the standard filter sets can be considered, and still avoid PWV induced effects.


\section{CONCLUSION}

In this study, we have presented the design, characterisation, and on-sky performance of SPIRIT, a novel near-infrared instrument tailored for the SPECULOOS survey to achieve optimal time-series photometric precision on late M and L type stars. By utilising an InGaAs-InP CMOS detector and a custom wide-pass \textit{zYJ} filter (0.81 -- 1.33~\unit{\um}), we have demonstrated improved photometric performance compared to the existing CCD-based instrument using the \textit{I+z'} filter (0.7 -- 1.1~\unit{\um}) at SSO for SPECULOOS' coolest targets.

The custom \textit{zYJ} filter significantly mitigated the impact of precipitable water vapour variability, a major source of red noise in near-infrared observations. Our on-sky tests confirmed a reduction in sensitivity to precipitable water vapour change by a factor of 15 compared to the existing instrumentation with the \textit{I+z'} filter.

Our noise models showed good agreement with the on-sky observations. We model that SPIRIT will outperform the existing instrumentation for targets with effective temperatures cooler than 2550~K. However, further investigation is necessary to understand the effects of bad pixels on photometry. 

The lessons learned from SPIRIT can be applied to future instruments, and areas for improvement include reducing read noise and optimising the field of view and pixel plate scale for the intended application. Furthermore, cooling the optical region to below $-$19~\si{\celsius} in front of the detector could potentially minimise thermal background flux-induced noise to sub-1~e$^-$/s/pixel levels.

Overall, this work demonstrates the feasibility and benefits of ground-based near-infrared photometry utilising InGaAs CMOS detectors. As detector technologies continue to improve, we anticipate that InGaAs-based instrumentation will become increasingly adopted for high-precision near-infrared time-series observations from the ground, a suitable alternative to HgCdTe-based instrumentation due to its lower cost and maintenance-free operations.

\newpage

\appendix    

\section{Quantum efficiency}\label{app:qe}

\begin{figure}[!h]
  \centerline{\includegraphics[width=0.8\textwidth]{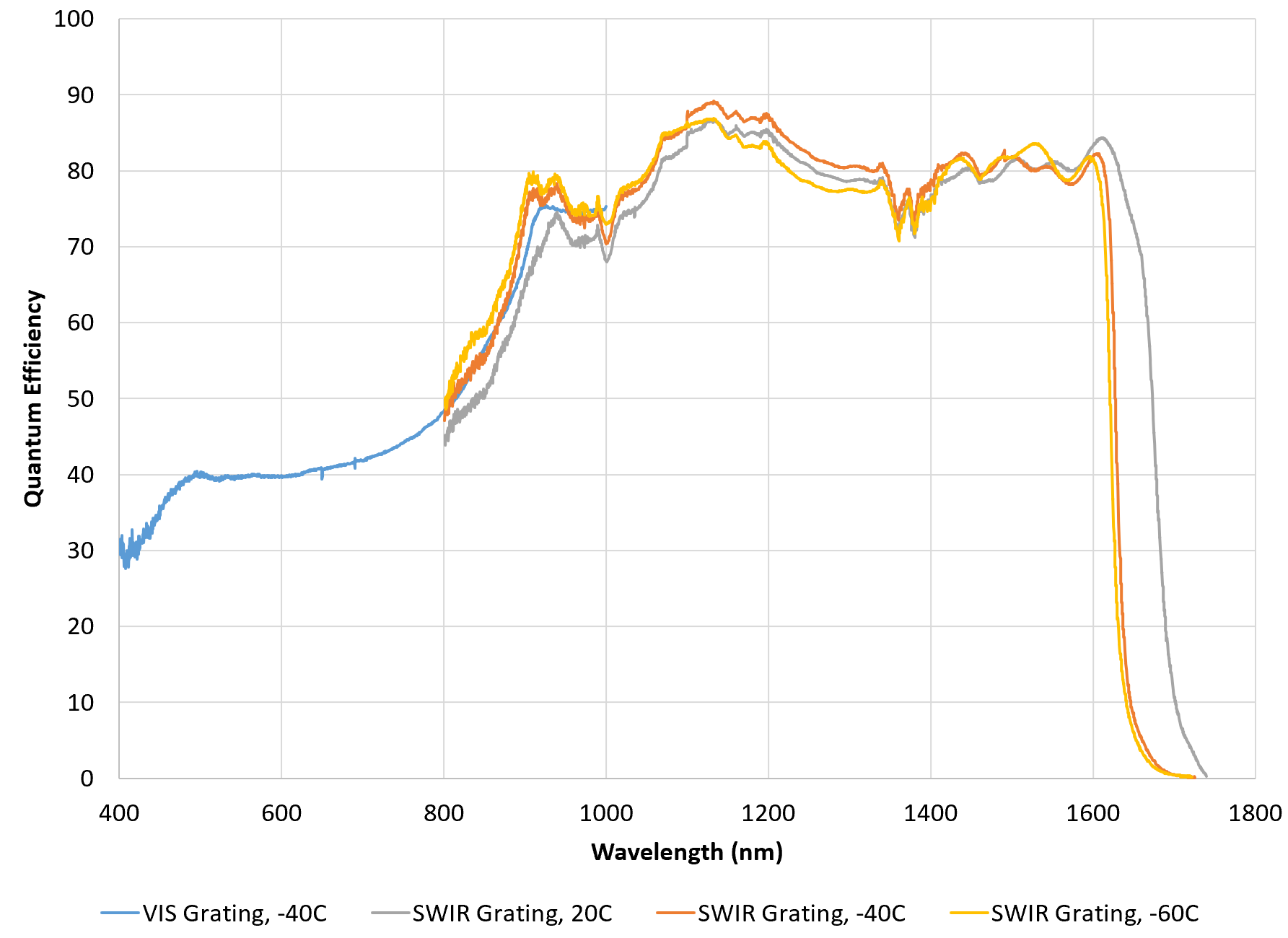}}
  %
  %
  \caption{Measured QE curve of the 1280SciCam by the camera manufacturer at three different FPA temperatures, +20~\si{\celsius}, $-$40~\si{\celsius}, and $-$60~\si{\celsius}. Figure credit: PIRT.}\label{f:real-qe}
\end{figure}

The high resolution quantum efficiency (QE) curve shown in Fig.~\ref{f:real-qe} was obtained by the manufacturer, which we did not have the resources to do in our lab. The QE was measured at three different focal plane array (FPA) temperatures (+20~\si{\celsius}, $-$40~\si{\celsius}, and $-$60~\si{\celsius}), for the complete camera build, i.e. not corrected for the fused silica window in front of the detector. The accuracy of the QE measurements was assumed to be within $\pm$10\% based on discussions with the manufacturer. One known feature of decreasing the FPA temperature was the QE cutoff shifting to lower wavelengths, which has been observed with other InGaAs based detectors \cite{seshadri2006characterization} and similarly with Si based detectors \cite{krishnamurthy2019precision}. At $-$60~\si{\celsius}, the cutoff (measured at a QE of 40\% to the closest nanometer on Fig.~\ref{f:real-qe}) was 1624~nm, whereas at +20~\si{\celsius} the cutoff was 54~nm higher at 1678~nm, and at $-$40~\si{\celsius} the cutoff was 1630~nm. A decreasing cutoff provides the benefit of lowering any environmental thermal influence on the detector, but at the disadvantage of lowering the sensitivity within the H band. A similar shifting effect was seen at $\sim$900~nm, the point where InP substrate removal permits QE into the visible domain. Without the InP substrate removal, the QE of an InGaAs detector would instead cut on at $\sim$900~nm \cite{seshadri2006characterization}. The sharp decline observed with our unit at this wavelength may suggest not all the InP substrate was removed. There is no conclusive evidence on the overall QE changing as a function of temperature given the error in the QE measurements. The mean QE across the \textit{zYJ} filter range was 77\%.

\newpage

\section{Linearity}\label{app:lin}

\begin{figure}[!h]
  \centerline{\includegraphics[width=0.8\textwidth]{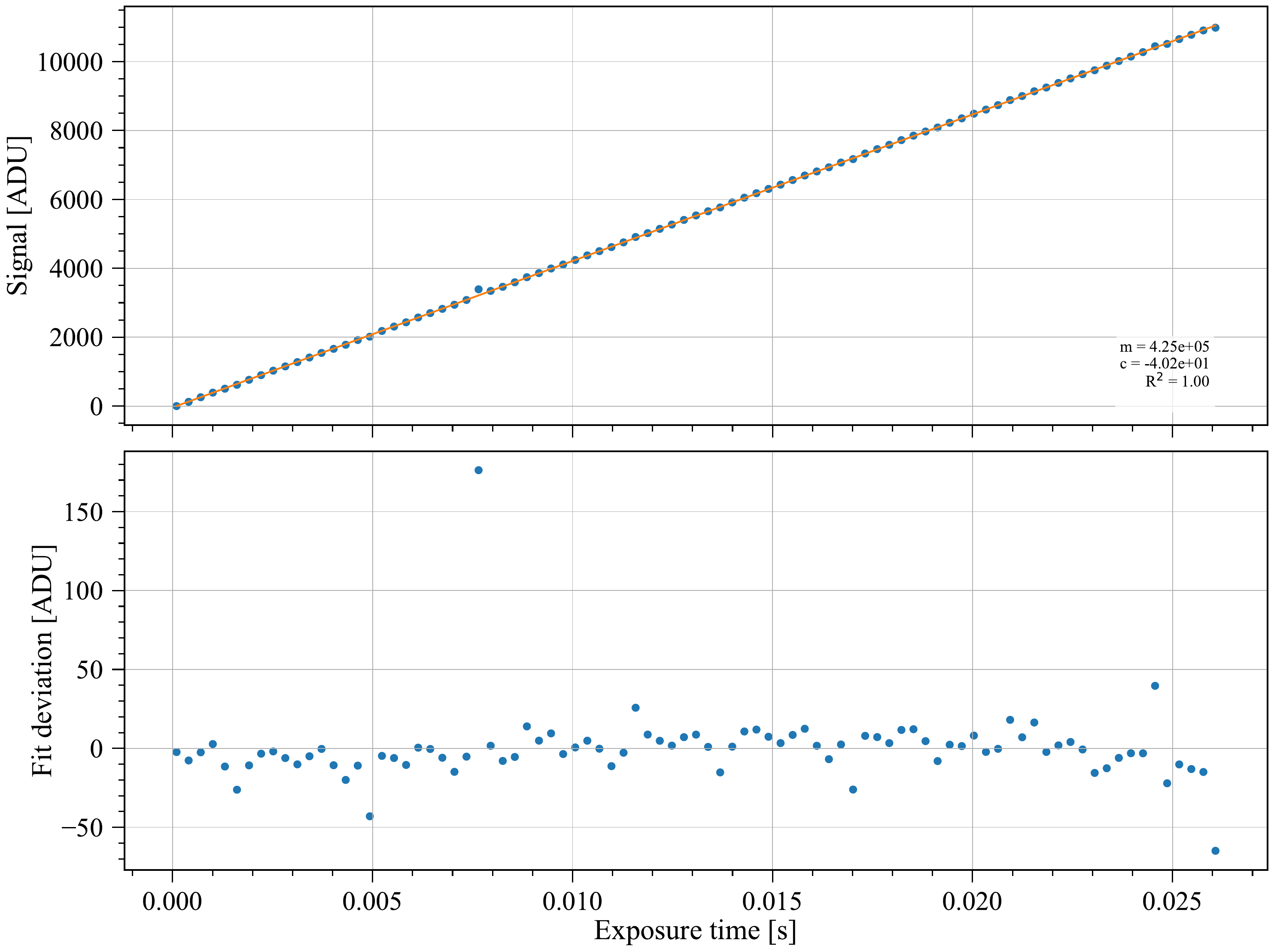}}
  %
  %
  \caption{Linearity measurement by the camera manufacturer. Top: Signal (mean of frames minus
  mean bias) against exposure time. Line of best ﬁt of blue data
  points, covering the range 0~s exposure to the exposure before saturation. Bottom: Deviation from line of best fit. Nonlinearity (\%) = [(Maximum Positive Deviation + Maximum Negative Deviation)/ Maximum Signal] x 100 = 0.95\%. This was calculated up to 11090~ADU, ignoring the single deviation at 0.0077~s -- assumed to be an artefact to that specific exposure setting.
  }\label{f:lin}
\end{figure}

\newpage

\section{Persistence}\label{app:per}

\begin{figure}[!h]
  \centerline{\includegraphics[width=0.8\textwidth]{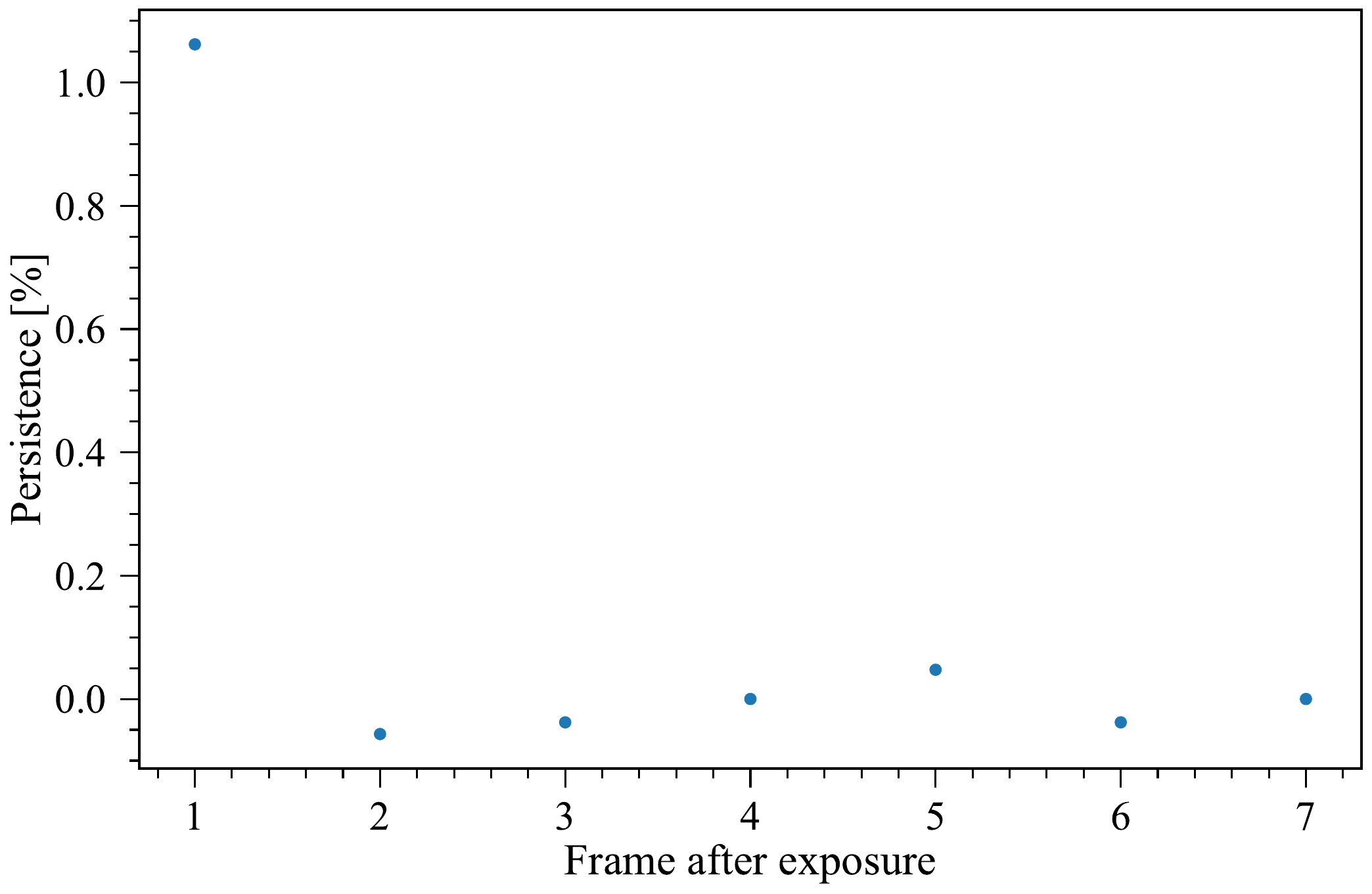}}
  %
  %
  \caption{Persistence measurements performed by the camera manufacturer, where
  a light source was triggered to be on for the one frame to fill $\sim$80\% of the well and then shut off for the next 7 subsequent frames. Frame time of 1.01s was used.
  }\label{f:per}
\end{figure}

\newpage

\section{\textit{zYJ} filter}\label{app:filter}

\begin{figure}[!h]
  \centerline{\includegraphics[width=0.8\textwidth]{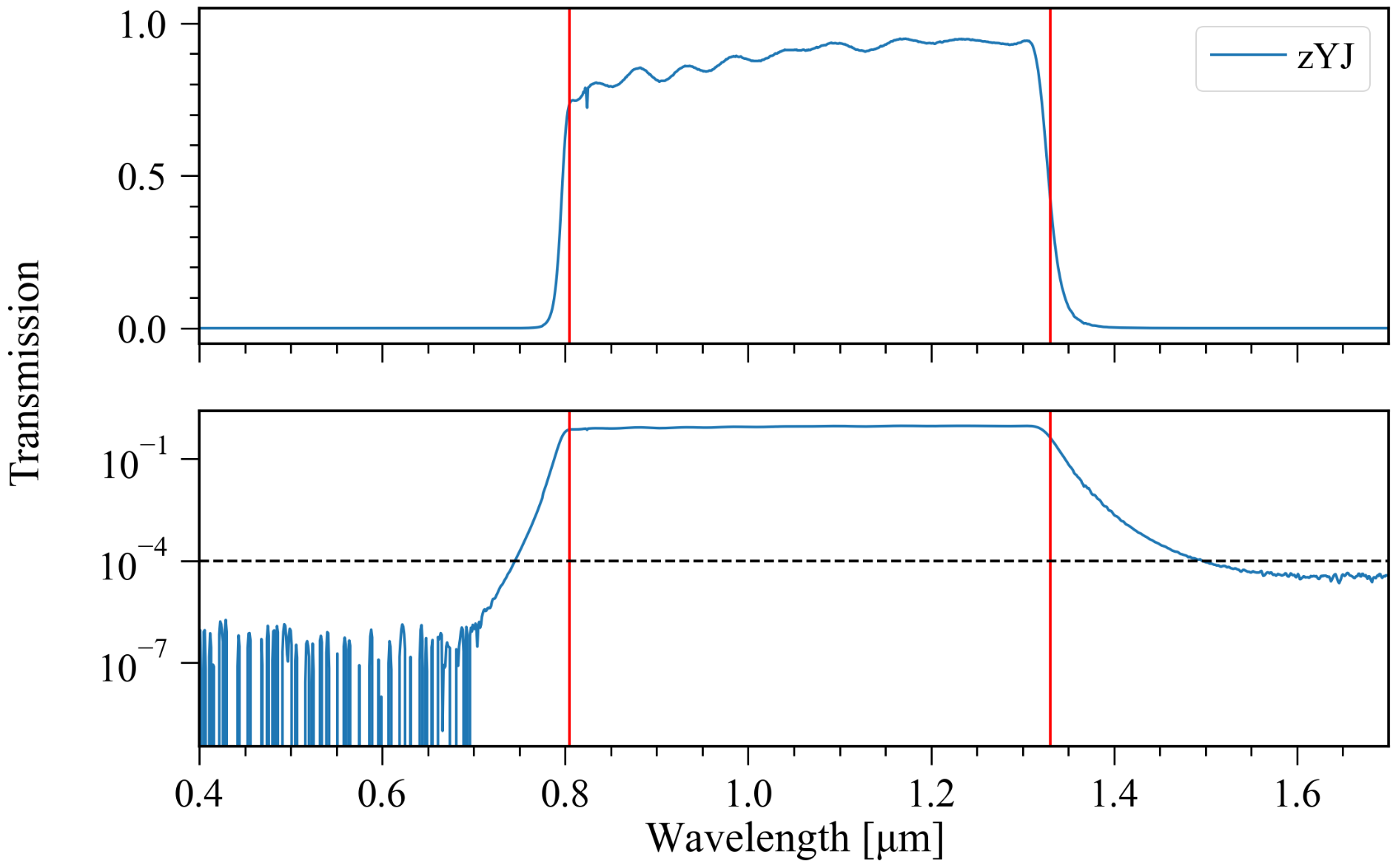}}
  %
  %
  \caption{Transmission data (in blue) of the accepted \textit{zYJ} filter set from Brinell Vision, with linear (top) and log-scale (bottom). Red vertical lines show the requested half power points from the specifications. In the log-scale plot, a dashed black line shows the requested OD4 blocking specification. The sharp dip at 0.82~\unit{\um} was assumed to be due to instrument anomalies as opposed to a real feature.
  }\label{f:filter-transmission}
\end{figure}

\newpage

\section{Modified liquid chiller}\label{app:chiller}

\begin{figure}[!h]
  \centerline{\includegraphics[width=0.8\textwidth]{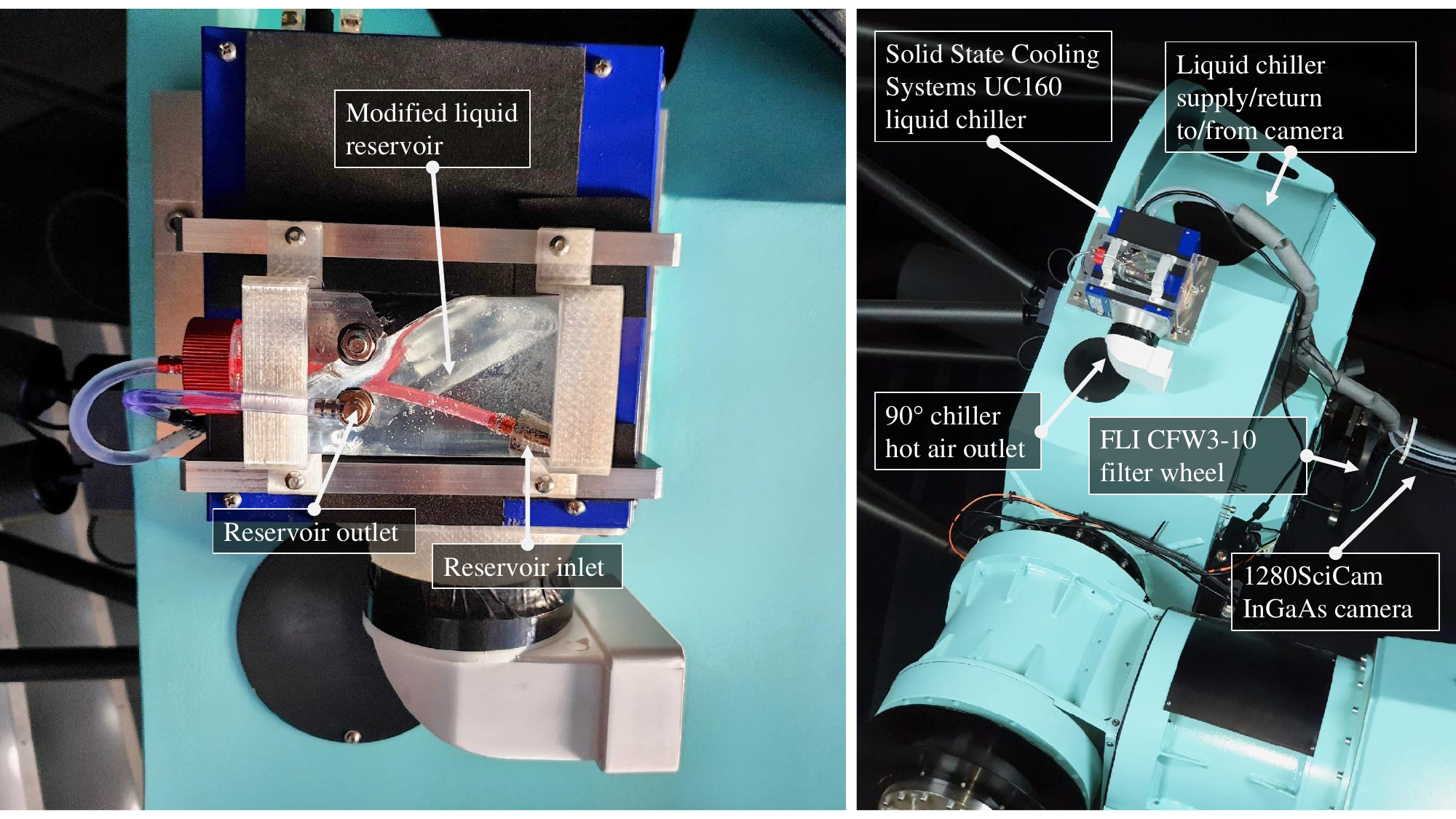}}
  %
  %
  \caption{Mounted modified liquid chiller on the side of the telescope, operating with a 30:70 mix of propylene glycol and water by volume respectively. Left: Closeup of mounted liquid chiller. Right: Relative positioning of the chiller on the telescope showing the tubing routed to the back of the camera, partially insulated with foam.
  }\label{f:chiller-closeup}
\end{figure}

\newpage

\acknowledgments 

The research leading to these results has received funding from the European Research Council (ERC) under the FP/2007--2013 ERC grant agreement n$^{\circ}$ 336480, and under the H2020 ERC grants agreements n$^{\circ}$ 679030 \& 803193; and from an Actions de Recherche Concert\'{e}e (ARC) grant, financed by the Wallonia--Brussels Federation. We also received funding from the Science and Technology Facilities Council (STFC; grants n$^\circ$ ST/S00193X/1, ST/00305/1, and ST/W000385/1). This work was also partially supported by a grant from the Simons Foundation (PI Queloz, grant number 327127), as well as by the MERAC foundation (PI Triaud), and the Balzan Prize foundation (PI Gillon). PPP acknowledges funding by the Engineering and Physical Sciences Research Council Centre for Doctoral Training in Sensor Technologies and Applications (EP/L015889/1). ED acknowledges support from the innovation and research Horizon 2020 program in the context of the Marie Sklodowska-Curie subvention 945298. This publication benefits from the support of the French Community of Belgium in the context of the FRIA Doctoral Grant awarded to MT. LD is an F.R.S.-FNRS Postdoctoral Researcher.
 
\bibliography{report} 
\bibliographystyle{spiebib} 

\end{document}